\documentclass[10pt,aps,twocolumn,prc,superscriptaddress,showpacs,nofootinbib,noshowkeys,floatfix,preprintnumbers]{revtex4}
\usepackage[dvips]{graphics,graphicx}
\usepackage{amsmath, amssymb}
\usepackage{multirow}
\usepackage{longtable}
\usepackage{color}
\usepackage[normalem]{ulem}  
\newcommand{\del}{\partial}

\renewcommand\sout{\bgroup \color{red} \ULdepth=-.5ex \ULset}

\begin{document}
\preprint{YITP-13-2}
\title{Net quark number probability distribution near the chiral crossover transition}
\date{\today}
\author{Kenji Morita}
\email{kmorita@yukawa.kyoto-u.ac.jp}
\altaffiliation[Present address: ]{Frankfurt Institute for Advanced
Studies, Ruth-Moufang-Str. 1, D-60438 Frankfurt am Main, Germany}
\affiliation{Yukawa Institute for Theoretical Physics, Kyoto University,
Kyoto 606-8502, Japan}
\affiliation{Extreme Matter Institute EMMI, GSI,
Planckstr. 1, D-64291 Darmstadt, Germany}
\author{Bengt Friman}
\affiliation{GSI, Helmholzzentrum f\"{u}r Schwerionenforschung,
Planckstr. 1, D-64291 Darmstadt, Germany}
\author{Krzysztof Redlich}
\affiliation{Institute of Theoretical Physics, University of Wroclaw,
PL-50204 Wroc\l aw, Poland}
\affiliation{Extreme Matter Institute EMMI, GSI,
Planckstr. 1, D-64291 Darmstadt, Germany}
\author{Vladimir Skokov}
\affiliation{Physics Department, Brookhaven National Laboratory, Upton, NY 11973, USA}
\begin{abstract}
 We investigate properties of the probability distribution of the net
 quark number near the chiral crossover transition in the  quark-meson model. The
 calculations are performed within the functional renormalization group
 {approach},  as well as in the mean-field approximation.  We find,   that there
 is a  substantial influence of the underlying  chiral phase transition
 on the properties of the probability distribution.  In particular, for a physical pion mass, the distribution which includes
 the effect of mesonic fluctuations, differs considerably from both,
 the mean-field and Skellam distributions. The latter is considered as a
 reference for a non-critical behavior. A characteristic feature of the
 net quark number probability distribution is that, in the vicinity of
 the chiral crossover transition in the $O(4)$ universality
 class,  it is narrower than the corresponding mean-field and  Skellam
 function. We study   the volume dependence  of the probability
 distribution, as well as the resulting cumulants, and discuss their
 approximate scaling properties.

\end{abstract}
\maketitle

\section{Introduction}

The structure of the QCD phase diagram  is one of the fundamental
problems addressed in both theoretical and
experimental
studies \cite{friman11:_cbm_physic_book,fukushima11:_phase_diagr_of_dense_qcd}.
At finite chemical potential, the existence of a critical point ({CP}) has
been conjectured, based on effective chiral models
\cite{asakawa89:_chiral_restor_at_finit_densit_and_temper} and
preliminary lattice QCD results \cite{lattice_CP}.
{The fluctuations of conserved charges,  have been proposed as a signature  for the
conjectured CP
\cite{hatta03:_proton_number_fluct_as_signal,stephanov98:_signat_of_tricr_point_in_qcd,ejiri06:_hadron_fluct_at_qcd_phase_trans,karsch11:_probin_freez_out_condit_in,stephanov09:_non_gauss_fluct_near_qcd_critic_point,stephanov11:_sign_of_kurtos_near_qcd_critic_point}.
However, in this paper we focus on the QCD phase transition at small net baryon densities.}

{It was conjectured by Pisarski and Wilczek
\cite{pisarski84:_remar_chiral_trans_in_chrom} that for massless light quarks,
the QCD phase transition is of the second order, belonging to
the $O(4)$ universality class.}
Current lattice QCD simulations at physical quark masses show,
that at vanishing baryon density the transition from a hadron gas to quark
matter is of the crossover type \cite{aoki06}.  This implies that the corresponding singularity in the
thermodynamic observables is shifted to complex values of the baryon
chemical potential~\cite{skokov11:_mappin}
{and the experimental signatures of the phase transition could be washed
out.}

{The conjecture of Pisarski and Wilczek is supported by} {recent lattice QCD
studies~\cite{ejiri09:_magnet_equat_of_state_in_flavor_qcd,Bazavov:2011nk}, which show that for physical values of the light quark masses, the  magnetic equation of state
of QCD  is consistent with the $O(4)$ scaling.}
This result has opened new opportunities to verify the QCD
phase boundary  experimentally by measuring fluctuations of conserved
charges
\cite{karsch11:_probin_freez_out_condit_in,braun-munzinger11:_net_proton_probab_distr_in,kaczmarek11:_phase_qcd,P_n_charge,bazavov12:_fluct_and_correl_of_net,Bazavov:2012vg}.
{Indeed,} based on the residual $O(4)$ criticality  and
the proximity of the chiral crossover transition to the freeze-out line in heavy
ion collisions, the characteristic modifications of the fluctuations of conserved charges  have been
proposed  {as a signature for}  the QCD phase boundary at small net baryon densities
\cite{ejiri06:_hadron_fluct_at_qcd_phase_trans,karsch11:_probin_freez_out_condit_in,braun-munzinger11:_net_proton_probab_distr_in,skokov10:_vacuum_fluct_and_therm_of_chiral_model,friman11:_fluct_as_probe_of_qcd,skokov11:_quark_number_fluct_in_polyak}.
It has been shown, that at the chiral crossover transition,  the higher order cumulants  of the
net baryon number and electric charge can be negative, {owing} to  the $O(4)$ scaling
\cite{skokov11:_quark_number_fluct_in_polyak,friman11:_fluct_as_probe_of_qcd}.

Such cumulants have recently been
explored in heavy ion collisions by  STAR Collaboration \cite{aggarwal10:_higher_momen_of_net_proton,luo_QM12}. The data show deviations
from the Skellam distribution, which  are qualitatively consistent with theoretical expectations based on the  $O(4)$ chiral
critical dynamics. We note, however, that the role of uncertainties associated with {the} event-by-event measurements of
fluctuations
remain to be clarified   \cite{Kitazawa:2011wh,Kitazawa:2012at,Bzdak:2012an,Bzdak:2012ab}.

Cumulants of net charges   have also been studied in first principle lattice QCD
calculations
\cite{allton05:_therm_of_two_flavor_qcd,bazavov12:_fluct_and_correl_of_net,borsanyi12:_fluct_of_conser_charg_at,Cheng:2008zh,Bazavov:2012vg},
as well as in effective chiral models
\cite{fukushima04:_chiral_polyak,sasaki07:_quark_number_fluct_in_chiral,sasaki07:_suscep_polyakov,stokic09:_kurtos_and_compr_near_chiral_trans,skokov10:_meson_fluct_and_therm_of,skokov10:_vacuum_fluct_and_therm_of_chiral_model,skokov11:_quark_number_fluct_in_polyak,friman11:_fluct_as_probe_of_qcd,asakawa09:_third_momen_of_conser_charg,herbst11:_phase_struc_of_polyak_quark,Schaefer-Wagner}. {Their
properties }  are consistent with general expectations based on the $O(4)$
scaling free energy.

Fluctuations of conserved charges are directly linked to the corresponding
probability distribution. Thus, the critical properties of cumulants of
conserved charges must {also be} reflected in  the probability
distribution.

Recently, the effects of the chiral phase
transition on the net baryon number probability distribution was examined within the
framework of {mean-field theory and the scaling theory} of phase transitions
\cite{morita12:_baryon_number_probab_distr_near}.
It was found,  that the critical behavior of the cumulants is a
consequence of the change of the corresponding probability distribution.

In the present work, we extend our previous studies to a
more realistic model.  We consider  the two-flavor
quark-meson model which is a low energy effective theory for the chiral
properties of QCD. The critical fluctuations,
are treated consistently by means of the functional renormalization group (FRG)
\cite{wetterich93:_exact_flow_equat,berges02:_non_pertur_renor_flow_in,delamotte07}.

In the quark-meson model the coupling of quarks to the Polyakov loop,
which is important for implementation of the  statistical confinement
properties, is neglected. However, within the FRG approach, the
$O(4)$ scaling behavior and thermodynamics near the CP are
well described by this model
\cite{stokic10:_frg_scaling,nakano10:_fluct_and_isent_near_chiral_critic_endpoin,Schaefer:2006ds,Kamikado}.


We show that there is a  substantial influence of the underlying  chiral
phase transition on the properties of the probability distribution.  In
particular, we find that for a physical pion mass, the distribution
which includes the effect of mesonic fluctuations,  differs considerably from both,  the mean-field and
Skellam distributions. The latter is considered as a reference
for a non-critical behavior. A characteristic feature of the
net quark number probability distribution  is   that in the vicinity of
{a chiral crossover transition of} the $O(4)$ universality class,  it is
narrower than the corresponding mean-field and  Skellam {distributions}.

This paper is organized as follows: in the next section, we introduce
the quark-meson model and its thermodynamic properties.
In Sec.~\ref{sec:pn}, we present results on the probability distribution
and  different cumulants of the net quark number. Section \ref{sec:summary} is
devoted to the concluding remarks.

\section{The thermodynamic potential in the Quark-Meson model}

We employ the  quark-meson model to explore the influence of the chiral
phase transition on the probability distribution of the net quark-number.

The quark-meson model is an effective realization of low energy QCD
in which the $O(4)$ chiral meson multiplet
$\phi=(\sigma,\vec{\pi})$  is
 coupled to quark fields. The
Lagrangian density is given by
\begin{gather}
 \mathcal{L}=\bar{q}[i\gamma_\mu \partial^\mu - g(\sigma + i\gamma_5
  \vec{\tau}\cdot\vec{\pi}) ]q +\frac{1}{2}(\partial_\mu
  \sigma)^2  +\frac{1}{2}(\partial_\mu \vec{\pi})^2 \nonumber\\-U(\sigma,\vec{\pi}),
\end{gather}
where $U(\sigma)$ denotes the meson potential,
\begin{equation}
U(\sigma,\vec{\pi}) = \frac{1}{2}m^2 \phi^2 + \frac{\lambda}{4}\phi^4 - h\sigma.
\end{equation}
For $m^2<0$ and $h=0$,  the $O(4)$ symmetry of the potential  is
spontaneously broken to $O(3)$, resulting in a  non-vanishing value of
the vacuum scalar condensate $\langle \sigma \rangle$ and a non-zero quark mass.
The last term,  $h=f_\pi m_\pi^2$,   breaks the chiral
symmetry explicitly and yields the nonzero pion mass.

We obtain the thermodynamics of  the quark-meson model  by computing the
thermodynamic potential within the FRG approach, as discussed in
Ref.~\cite{stokic10:_frg_scaling}.
Following   \cite{wetterich93:_exact_flow_equat}, we consider a scale
dependent effective action in the local potential approximation.
Thereby, we neglect the wave function renormalization and the flow of the Yukawa
coupling. Using the so-called optimized cutoff functions, one obtains the
evolution equation for the scale dependent thermodynamic potential
density \cite{stokic10:_frg_scaling} with the reduced field variable   $\rho = (\sigma^2 +
\vec{\pi}^2)/2$,
\begin{widetext}
\begin{gather}
 \partial_k \Omega_k(\rho)=\frac{k^4}{12\pi^2}\left[
 \frac{3}{E_\pi}\left\{1+2n_B(E_\pi)\right\}
	+\frac{1}{E_\sigma}\left\{1+2n_B(E_\sigma)\right\}
 -\frac{2\nu_q}{E_q}\left\{
 1-n_F(E_q^+)-n_F(E_q^-)\right\}
 \right]\label{eq:floweq},
\end{gather}
where $n_B$ and $n_F$ are the Bose and the Fermi distribution functions,
respectively and  $\nu_q=2N_c N_f =12$ is the quark degeneracy. The
single particle energies of pion, sigma meson and quark/antiquark are given by
 \begin{align}
    E_\pi=\sqrt{k^2 + \bar{\Omega}^\prime_k}, ~~
    E_\sigma = \sqrt{k^2 + \bar{\Omega}^\prime_k + 2\rho
    \bar{\Omega}^{\prime\prime}_k}, ~~
    E_q^\pm &= \sqrt{k^2 + 2g^2 \rho}\pm \mu ,
 \end{align}
where $\bar{\Omega}_k^{\prime}$ and
$\bar{\Omega}_k^{\prime\prime}$ denote the first and the second derivatives of
$\bar{\Omega}_k=\Omega_k+h\sqrt{2\rho_k}$, with respect to $\rho$.
The flow equation \eqref{eq:floweq} is solved by using the Taylor
expansion method. Expanding the potential up to the third order in
$\rho$ around the potential minimum $\rho_k$,
\end{widetext}
\begin{equation}
  \Omega_k(\rho) = \sum_{n=0}^{3} \frac{a_n(k)}{n!}(\rho-\rho_k)^n,
\end{equation}
and using Eq.~\eqref{eq:floweq},  one finds the flow equations for the coefficients
$a_n(k)$ and $\rho_k$,
\begin{eqnarray}
d_ka_{0,k}&=&\frac{c}{\sqrt{2\rho_k}}\,d_k\rho_k+\del_k\Omega_k,
\nonumber\\
d_k\rho_k&=&-\frac{1}{\left(c/(2\rho_k)^{3/2}+a_{2,k}\right)}
\del_k\Omega'_k,\label{eq:FRG-Taylor}
\\
d_ka_{2,k}&=&a_{3,k}\, d_k\rho_k+\del_k\Omega''_k,
\nonumber\\
d_ka_{3,k}&=&\del_k\Omega'''_k,\nonumber
\end{eqnarray}
where  $d_k=d/dk$.
The flow equations are solved numerically starting at the ultraviolet
cutoff scale $\Lambda=1.0$ GeV \cite{stokic10:_frg_scaling}.
We eliminate $a_1$ by means of the scale independent relation $h =
a_1(k)\sqrt{2\rho_k}$.

There are  four initial conditions for the flow equations, which are fixed
at the scale $k=\Lambda$. Within this scheme, the initial value of $a_{0}$
is just  a constant shift of thermodynamic potential $\Omega$.
We note, however, that such a
cutoff at a finite momentum leads to an unphysical behavior
of thermodynamic quantities at high temperatures. This problem can be amended by
accounting for the $\mu-$ and $T-$dependent contribution of the momenta beyond the cutoff
scale \cite{Braun_PRD70,skokov10:_meson_fluct_and_therm_of}. Following
Ref.~\cite{Braun_PRD70}, we include the high-momentum contribution approximately by
using the flow equation for non-interacting massless quarks and gluons,
\begin{align}
 \partial_k \Omega_k^\Lambda(T,\mu) &=
  \frac{k^3}{12\pi^2}\{2(N_c^2-1)[1+2n_B(k)] \nonumber\\
 & \, - \nu_q [1-n_F(k^+)-n_F(k^-)] \}.\label{eq:flow_SB}
\end{align}
By integrating the flow equation \eqref{eq:flow_SB} from $k=\infty$ to $k=\Lambda$,
we obtain $\Omega^\Lambda(T,\mu)$ which is then used as initial condition $a_0(\Lambda)$
for the solution of the flow equations (\ref{eq:FRG-Taylor}).

We set $a_3(\Lambda)=0$ and fix $\rho_{k=\Lambda}$  and $a_2(\Lambda)$
by requiring that in vacuum  the pion  $m_\pi=135$ MeV and the sigma
 $m_\sigma= 640$ MeV masses  are reproduced.  The strength of the
Yukawa coupling, $g=3.2$, is fixed by the constituent quark mass,
$M_q(T=\mu=0) = g\sigma_{k=0}(T=\mu=0)  = 300$ MeV with
$\sigma_{k=0}(T=\mu=0)$ $=$ $f_\pi =93$ MeV. The full thermodynamic
potential density of the quark-meson model $\Omega(T,\mu)$, which
includes thermal and quantum fluctuations of the meson and quark fields,
is then obtained from $\Omega(T,\mu)=\lim_{k\to 0}\Omega_k$,  where
$\Omega_{k}$ is the solution  of the flow
equation~\eqref{eq:floweq},\eqref{eq:FRG-Taylor}.

By ignoring the mesonic contribution in the flow equation
\eqref{eq:floweq},  we obtain the effective potential corresponding to
the mean-field approximation, with  a finite cutoff $\Lambda$. The fermionic vacuum
fluctuations, which are included  in the mean-field (MF) potential,  are
 necessary  to reproduce the second order phase transition at
 vanishing $\mu$ in the chiral limit \cite{nakano10:_fluct_and_isent_near_chiral_critic_endpoin,
 skokov10:_vacuum_fluct_and_therm_of_chiral_model}.
 The vacuum contribution must  be renormalized to remove the
artificial  cutoff dependence
 \cite{skokov10:_vacuum_fluct_and_therm_of_chiral_model}.
Then, in the mean-field approximation,  the thermodynamic potential is
given by,
\begin{align}
 \Omega_{\text{MF}}(\langle \sigma \rangle;T,\mu) &=
  U(\langle \sigma \rangle,\vec{\pi}=0)-\frac{\nu_q}{16\pi^2}M_q^4
  \ln\left(\frac{M_q}{M}\right) \nonumber \\
  -\nu_q T \int\! \frac{d^3 p}{(2\pi)^3} &
  \left[ \ln(1+e^{-(E_q-\mu)/T}) + \ln(1+e^{-(E_q+\mu)/T}) \right],
\end{align}
where $M$ is  an arbitrary renormalization scale parameter,  $M_q=g \langle \sigma \rangle$ and the expectation value $\langle
\sigma \rangle$ is determined by
 the solution of the gap equation
$\partial \Omega_{\text{MF}}/\partial \langle \sigma \rangle = 0$.

 \begin{figure*}[ht]
  \includegraphics[width=0.45\textwidth]{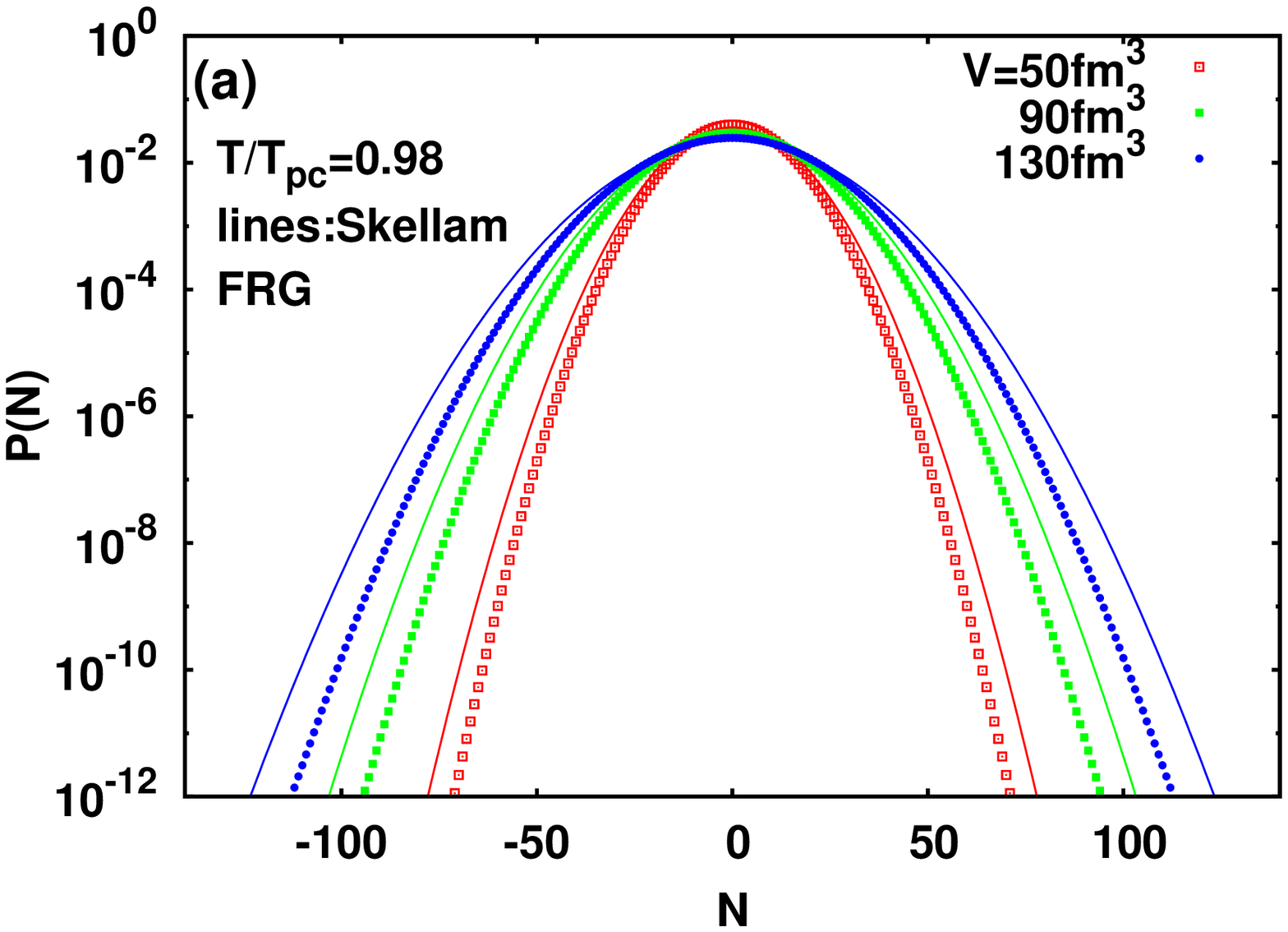}
  \includegraphics[width=0.45\textwidth]{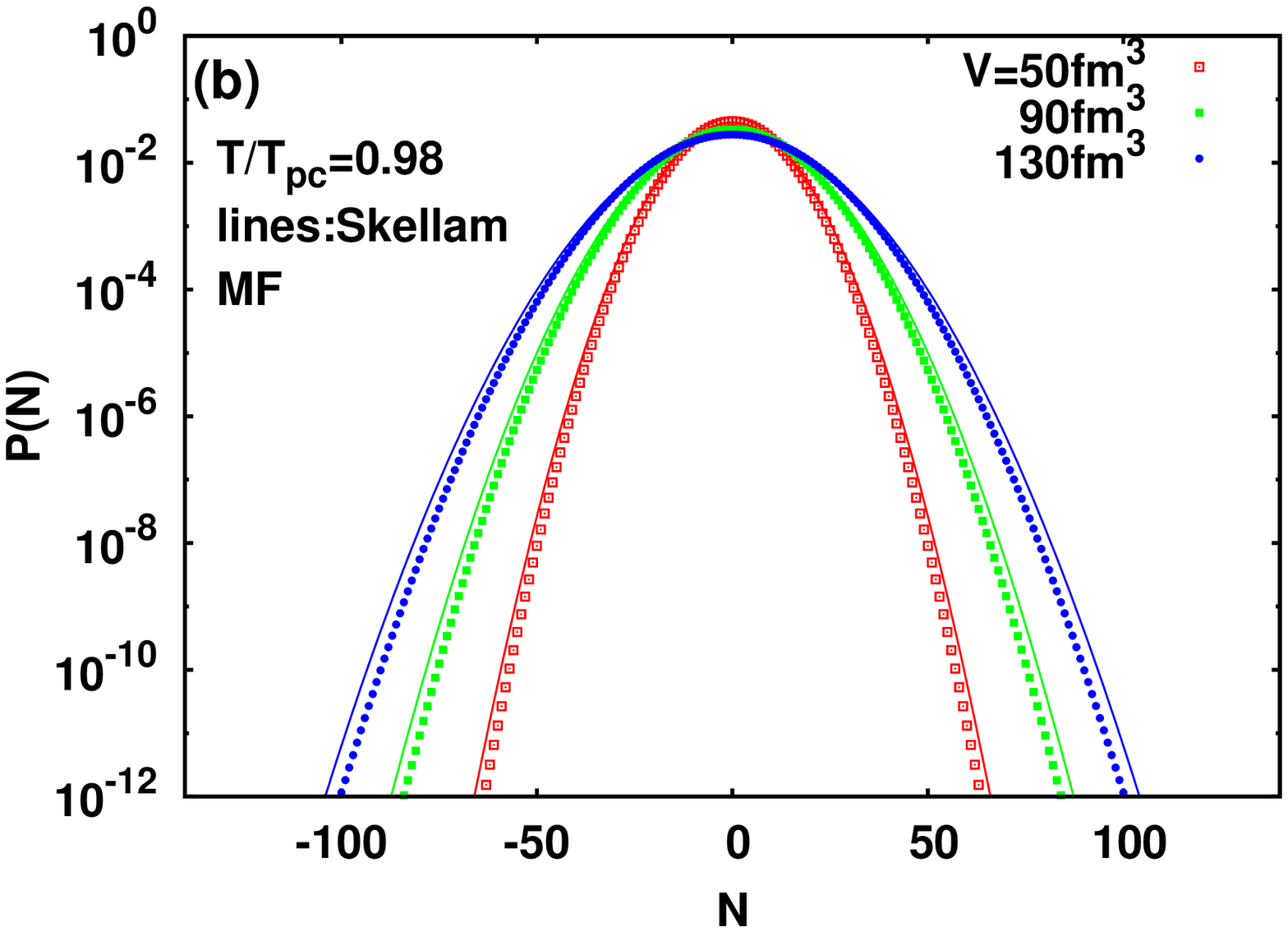}
  \caption{(Color online) Probability distributions of the net quark number just below
  the pseudocritical temperature $T_{pc}$ (a) in the FRG approach and (b)
  in the mean-field (MF) approximation, compared with the corresponding Skellam
  distributions. The dots are  the model results  for different
  volumes while the lines show the Skellam distribution.}
  \label{fig:pn_tc}
 \end{figure*}

\section{Probability distribution of the net  quark  number}
\label{sec:pn}
The thermodynamical potentials for the quark-meson model, derived in the previous
section,  can be used to assess the influence of the underlying chiral phase
transition on observables. In the following,
we focus on the  probability distribution of the net quark number, $P(N)$
and the corresponding cumulants $c_n(T,\mu)$.

\subsection{General properties of $P(N)$}
\label{subsec:pn}

We consider a thermodynamic system described by the grand canonical
ensemble at temperature $T$ in  a subvolume $V$. We introduce a
chemical potential $\mu$ which is used to tune the corresponding average net charge.
For the net quark number $N=N_q-N_{\bar q}$,
the  probability distribution $P(N)$ to find
 the net charge  $N$ in  volume $V$ is given by
\begin{equation}
P(N;T,\mu,V) = \frac{Z(T,N,V)}{\mathcal{Z}(T,\mu,V)}e^{\mu N/T},\label{eq:pn}
\end{equation}
where $Z(T,N,V)$ is the canonical and $\mathcal{Z}(T,\mu,V)$ the
grand-canonical partition function.
The normalization of the probability distribution,  $\sum_{N=-\infty}^{\infty} P(N)=1$, follows
from the fugacity expansion of the grand canonical partition function
\begin{equation}
 \mathcal{Z}(T,\lambda,V) = \sum_{N}\lambda^N Z(T,N,V)\label{eq:fug_exp},
 \end{equation}
where $\lambda=e^{\mu/T}$. Consequently, all essential information on $P(N)$ is
contained in the canonical partition function. Since the fugacity expansion is the  Laurent series in
the complex $\lambda$ plane, with coefficients given by the canonical partition function,
the latter can be obtained by performing the contour integral,
\begin{equation}
 Z(T,N,V) = \frac{1}{2\pi i }\oint_C d\lambda
  \frac{\mathcal{Z}(T,\lambda,V)}{\lambda^{N+1}}.\label{eq:cano}
\end{equation}
Thus, to compute the canonical partition function, one needs to know the
analytic structure of $\mathcal{Z}(T,\lambda,V)$ in the complex
$\lambda$ plane and to choose an appropriate integration
contour.\footnote{ In a finite system, the grand partition function
$\mathcal{Z}$ has Yang-Lee zeroes at complex values of $\lambda$. In the
thermodynamic limit, the zeroes join into cuts. }
In chiral effective models, the  structure of the singularities
associated with the chiral phase transition has been discussed in
Refs.~\cite{stephanov06:_qcd_critic_point_and_compl,skokov11:_mappin}.
In the broken phase, $T < T_c(\mu)$, there are no singularities on the unit
circle $\lambda=e^{i\theta}$ in the range $0 \leq \theta < 2\pi$.
Consequently, the canonical partition function
is obtained from \cite{Statmodelreview_QGP3,hagedorn85:_statis},
\begin{equation}
 Z(T,N,V) = \frac{1}{2\pi}\int_{0}^{2\pi}d\theta e^{-i\theta N}
  \mathcal{Z}(T,\theta,V)\label{eq:cano_theta},
\end{equation}
where $\theta=\mu_I/ T$ and $\mu_I$ is  the imaginary chemical potential.

The above equation links the grand canonical partition function in a finite volume $V$, at
imaginary chemical potential,  to the thermodynamics at fixed
net charge $N$. At imaginary $\mu_I$, the QCD  partition function
exhibits Roberge-Weiss periodicity,
$\mathcal{Z}(T,\theta+2\pi/3,V) = \mathcal{Z}(T,\theta,V)$
 \cite{roberge86:_gauge_qcd}.
The Polyakov loop extended effective chiral models reproduce the
Roberge-Weiss periodicity
\cite{sakai08:_polyak_nambu_jona_lasin,morita11:_probin_decon_in_chiral_effec,morita11:_role_of_meson_fluct_in}.
In the quark-meson model employed here,
the
period of the partition function in imaginary chemical potential  $\theta$ is $2\pi$.

In the present work, we follow
Ref.~\cite{morita12:_baryon_number_probab_distr_near} and compute the
canonical partition function
and the corresponding probability distribution using
Eq.~\eqref{eq:cano_theta}.
The thermodynamic potential  density  $\Omega=-(T/V)\log \mathcal{Z}$
is obtained in the quark-meson model within the FRG approach  as well as in
the mean-field approximation.
%
%

Because of the oscillatory nature of the integrand, a numerical
integration of Eq.~\eqref{eq:cano_theta} becomes unreliable for
large $|N|$. The numerical quadrature employed here yields accurate results up to
values of $|N|$ corresponding to $P(N;\mu=0)\sim 10^{-12}$,
independent of the volume, temperature and other parameters. As we show, the
achieved precision is sufficient for studying the influence of chiral
criticality on the properties of the net quark number probability
distribution.

\begin{figure*}[ht]
  \includegraphics[width=0.45\textwidth]{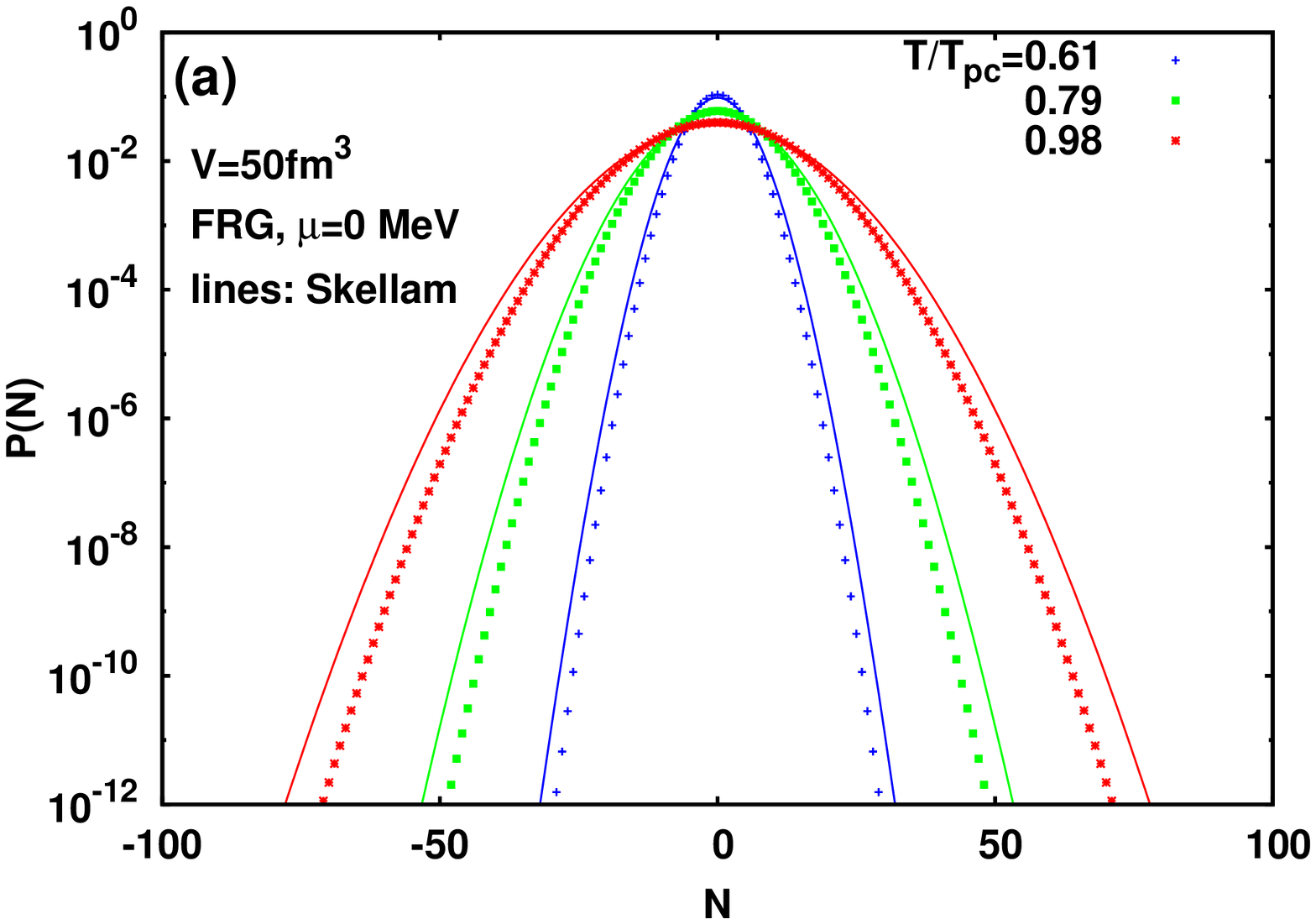}
  \includegraphics[width=0.45\textwidth]{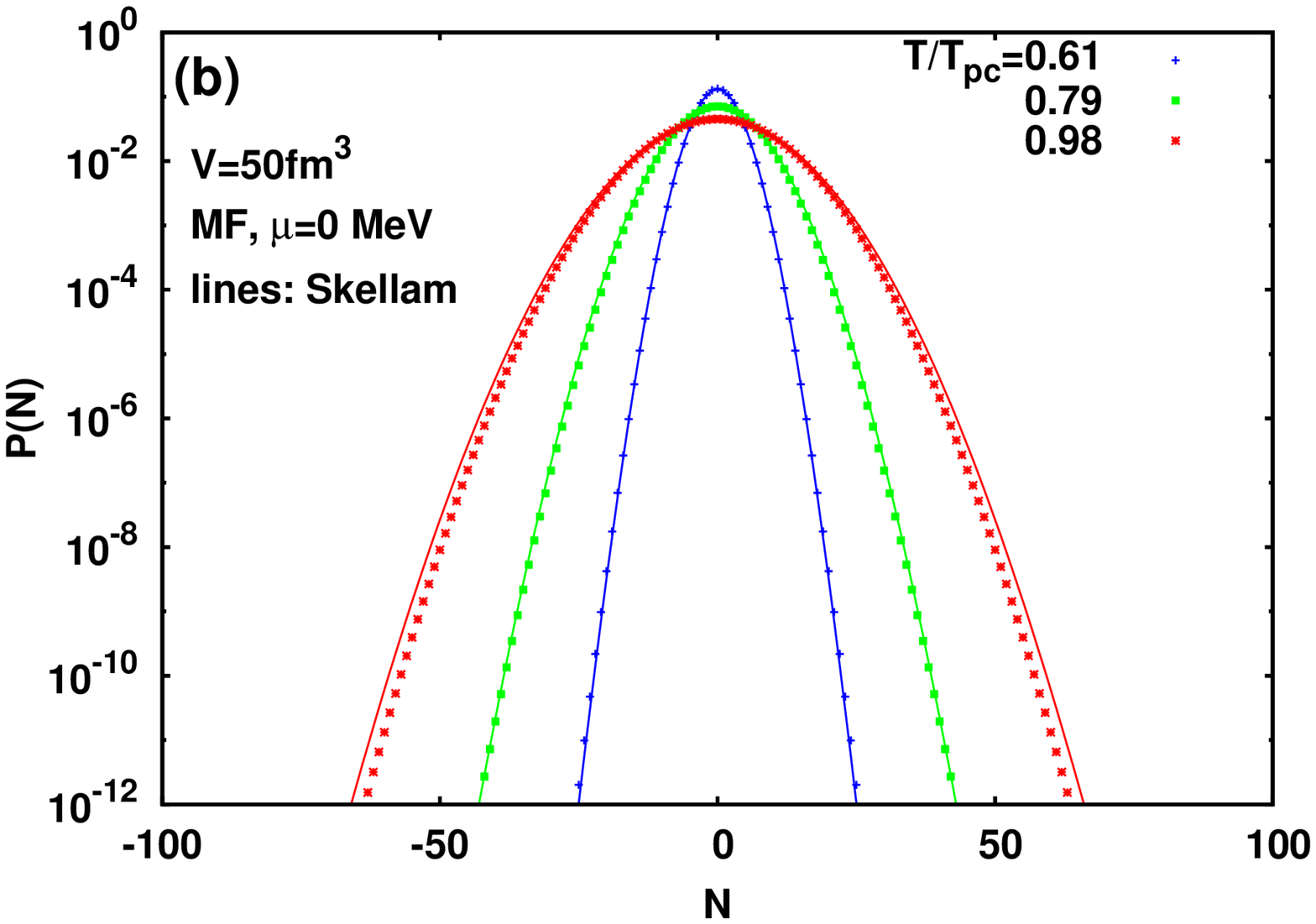}
  \caption{(Color online) Probability distribution for different
 temperatures (a) in the  FRG approach  and (b) in the mean-field (MF) approximation.
 The dots are  the model results  while the lines show the corresponding
 Skellam distribution (see text). }
  \label{fig:pn_Tdep}
 \end{figure*}

\subsection{Net quark number probability distribution near the chiral phase transition}

At vanishing and moderate values of $\mu$,   the
quark-meson model in the chiral limit exhibits the  second order phase transition,
belonging to the $O(4)$ universality class \cite{stokic10:_frg_scaling}.
For a physical pion mass, the chiral symmetry is explicitly broken and
the transition is of the crossover type.
Nevertheless, remnants of $O(4)$ criticality remain  in various
observables \cite{karsch11:_probin_freez_out_condit_in}. Thus,  also
the probability distribution of the net quark number is expected to exhibit
characteristic features reflecting the critical behavior of the underlying $O(4)$ transition.
We note,  that long range critical correlations can be unfolded only in
a sufficiently large sub-volume $V$,  and  close to
the pseudocritical temperature, $T_{pc}$.

To unravel the influence of chiral transition on the probability
distribution $P(N)$,
one needs  to establish a reference distribution, which does not include the effect
of critical fluctuations. At low temperatures, $T \ll T_{pc}$,
the thermodynamic potential
is expected to  be well described  as a quasi-ideal quark gas with a
dynamically generated temperature-dependent mass. Consequently, at fixed
$T$ and $V$,  the natural reference for $P(N)$ is the probability
distribution of an ideal gas of particles and antiparticles, i.e. the
Skellam distribution
\cite{braun-munzinger11:_net_proton_probab_distr_in}. The Skellam
distribution is  then  determined entirely by the mean number of quarks
$b= \langle N_q \rangle$ and antiquarks $\bar b= \langle N_{\bar q} \rangle$,
\begin{equation}
P(N) = \left( \frac{b}{\bar{b}} \right)^{N/2} I_N(2\sqrt{b\bar{b}}) e^{-(b+\bar{b})},\label{eq:pn_skellam}
\end{equation}
where  $I_N(x)$ is the modified Bessel function of the first kind.
The mean number of quarks $b$ is calculated as for an ideal
gas of constituent quarks with a dynamically generated mass
\begin{equation}
 b = \frac{\nu_q V}{2\pi^2}\int_{0}^{\infty}dk k^2 n_F(E_k;T,\mu),\label{eq:skellam_baryonnum}
\end{equation}
where $n_F$ is the Fermi distribution function and $E_k =
\sqrt{k^2+M_q^2}$ is the energy of a particle with momentum $k$. The
$\bar{b}$ is obtained from \eqref{eq:skellam_baryonnum}  by replacing $\mu\rightarrow -\mu$.

In the MF approximation
$M_q=g\langle \sigma \rangle$, while in the FRG approach,  we use the
scale dependent mass $M_{q,k}=g \sigma_{k}$,  which
for $k \leq \Lambda$ is obtained from the flow equation
\eqref{eq:floweq}.  For $\Lambda < k < \infty$  the  quarks are
assumed to  be massless and non-interacting, as discussed above.

 Figure~\ref{fig:pn_tc}   shows  the probability distributions of the net
quark number obtained in the quark-meson model, within the FRG
and in the MF approximation  at
vanishing chemical potential and near the pseudocritical
temperature $T_{pc}$. The  $T_{pc}$ corresponds to the peak
position of the chiral susceptibility, which in the FRG and MF calculations
are located at 214 MeV  and 190 MeV, respectively.

The probability distributions in Fig.~\ref{fig:pn_tc} are
calculated for different volumes and at fixed  temperature.
There is a clear change of  distributions with the volume  as a consequence of
a linear dependence  of the variance on  $V$.
%
%
The probability distribution also changes rapidly as the
temperature is lowered from the pseudocritical point. As shown in
Fig.~\ref{fig:pn_Tdep}, at a given volume,  the distributions are
narrower at smaller temperatures. This is due to   growing  mass of
quarks,  which together with decreasing temperature,  imply
decreasing  mean number of quarks and antiquarks, and consequently the
width of the distribution.

The influence of  the criticality on the net-quark number
probability distribution,  and the differences between the  MF and the
FRG dynamics, are particularly transparent when comparing the
results with the non-critical Skellam function.

Figures \ref{fig:pn_tc} and \ref{fig:pn_Tdep} show,  that near $T_{pc}$,
both the MF and the FRG distributions are narrower than their Skellam
counterparts. Such reduction of the width of the probability
distribution compared to the Skellam distribution was already seen in
results obtained previously in the Landau theory of phase transitions
when critical fluctuation is included to give negative higher order
cumulants~\cite{morita12:_baryon_number_probab_distr_near}.
The deviations from Skellam distribution are stronger when mesonic
fluctuations are included, i.e.\ in the FRG approach. Except for the
highest temperature, $P(N)$ in the MF approximation coincides with the
Skellam distribution, while in the FRG calulations, the two
distributions differ at all temperatures. The increasing difference
between the FRG and Skellam distributions, as the temperature
approaches $T_{pc}$, reflects the growing influence of
mesonic fluctuaions leading to the $O(4)$ criticality.

\begin{figure}[!t]
 \includegraphics[width=3.375in]{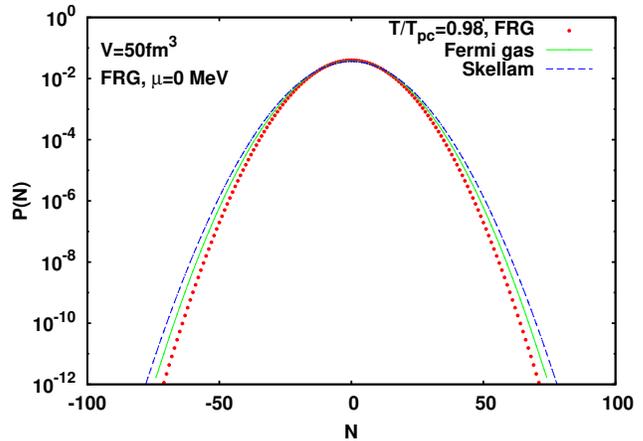}
 \caption{(Color online) The probability distribution obtained within
 the FRG approach (red, circles), the Skellam distribution (blue,
 dashed), and that of a free Fermi gas (green, solid). All distributions
 correspond to the same average quark and antiquark numbers, $b$ and
 $\bar{b}$, given by (\ref{eq:skellam_baryonnum}).}
 \label{fig:fermi}
\end{figure}

\begin{figure}[!t]
  \includegraphics[width=3.375in]{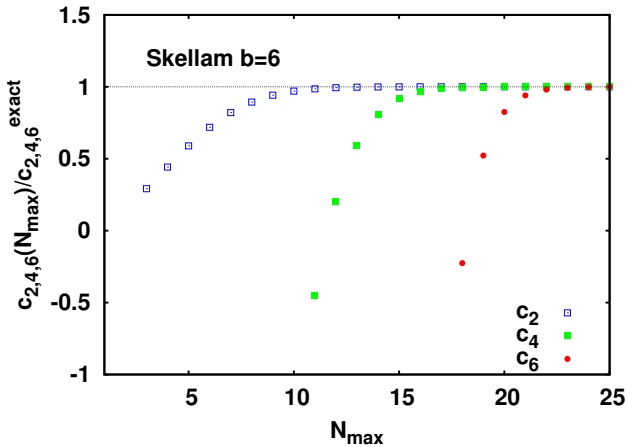}
  \caption{(Color online) The ratios of the cumulants $c_n$ $(n=2,4,6)$ of the net charge
 obtained from the Skellam distribution with the mean multiplicities
 $b=\bar b=6.0$ using
 Eqs.~\eqref{nmax}-\eqref{moments} for a given value of $N_{\text{max}}$ to their  corresponding exact  values.}
  \label{fig:comps}
 \end{figure}



\begin{figure*}[ht]
  \includegraphics[width=0.32\textwidth]{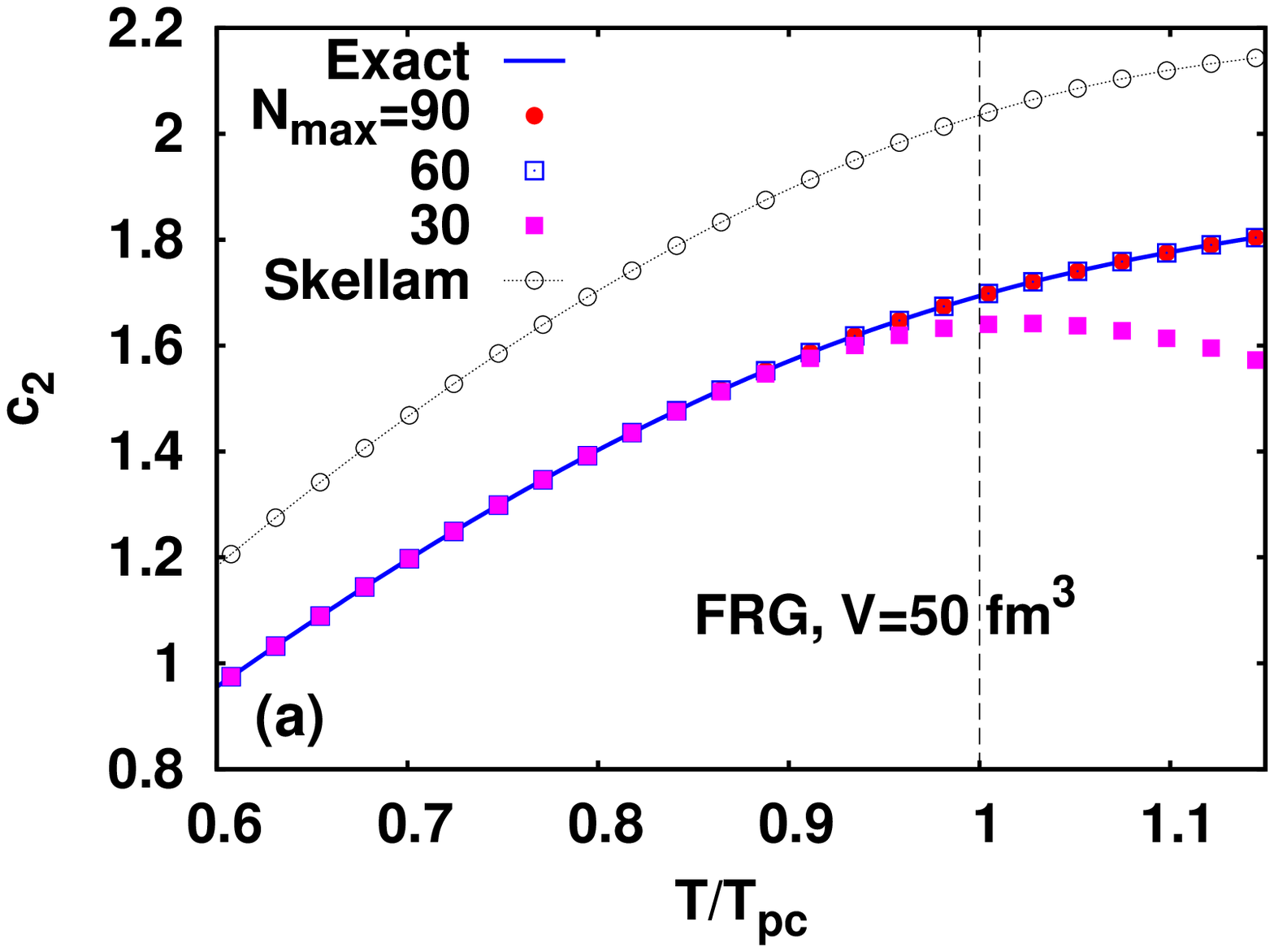}
  \includegraphics[width=0.32\textwidth]{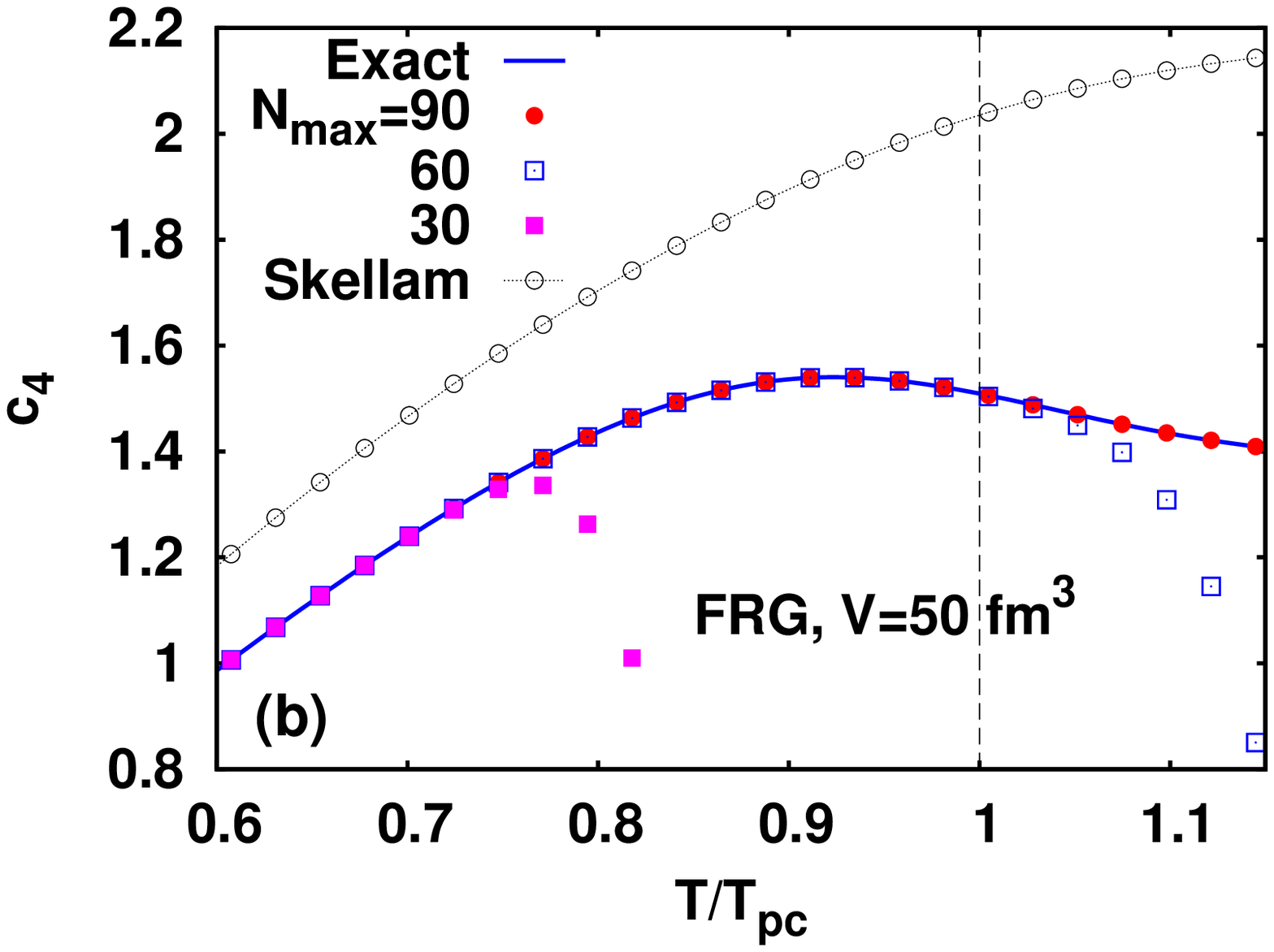}
\includegraphics[width=0.32\textwidth]{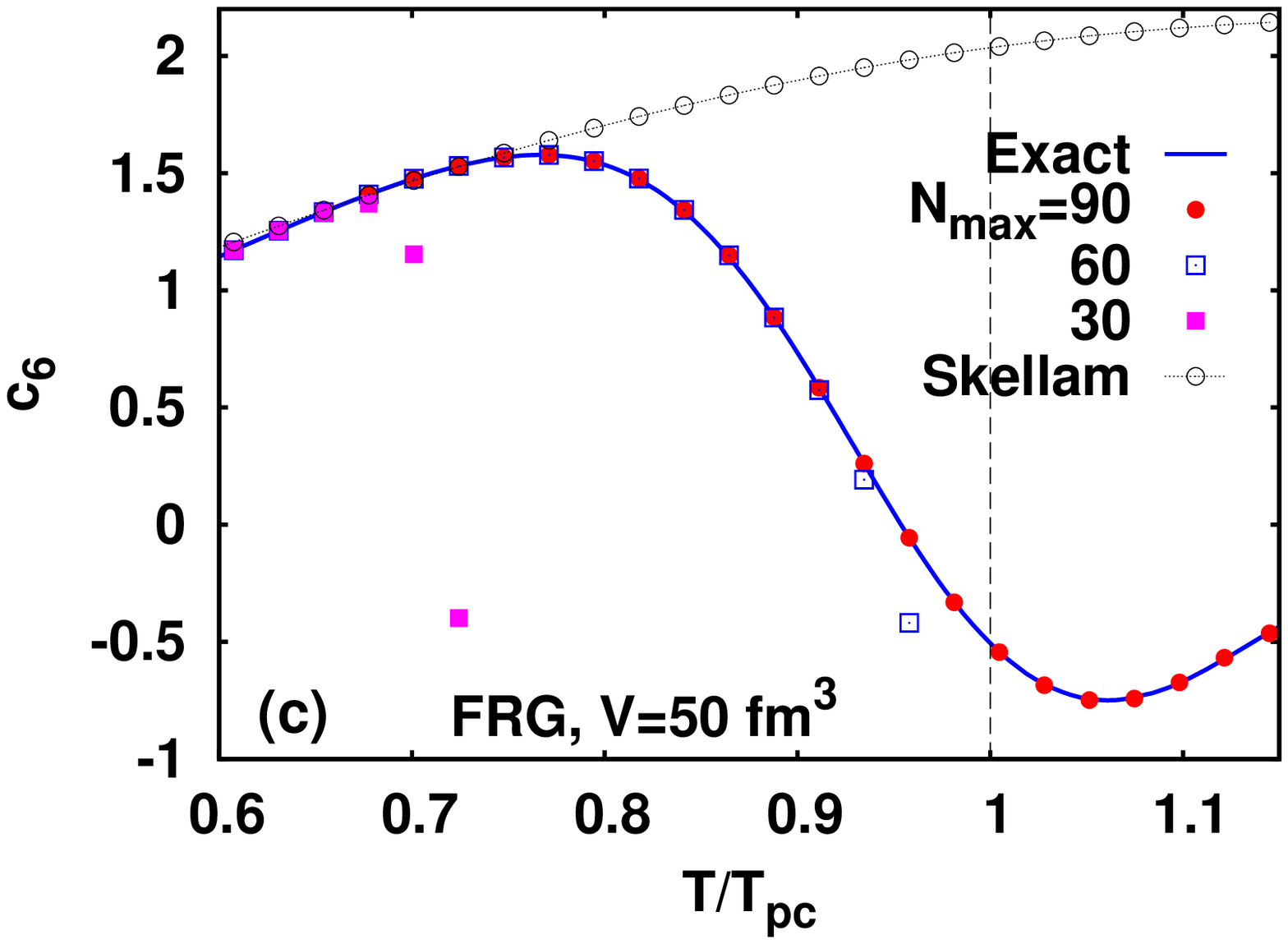}
  \caption{(Color online) Cumulants $c_n$ $(n=2,4,6)$  obtained in the quark-meson model
 within the FRG approach  using Eq.~\eqref{cumulants} (full line)  and
 Eqs.~(\ref{nmax}-\ref{moments}) (points)  for several values
 of $N_{\text{max}}$. Results for the corresponding Skellam distribution
 are also indicated by the lines with open
 circles.}
  \label{mom24}
 \end{figure*}

\begin{figure*}[ht]
 \includegraphics[width=0.32\textwidth]{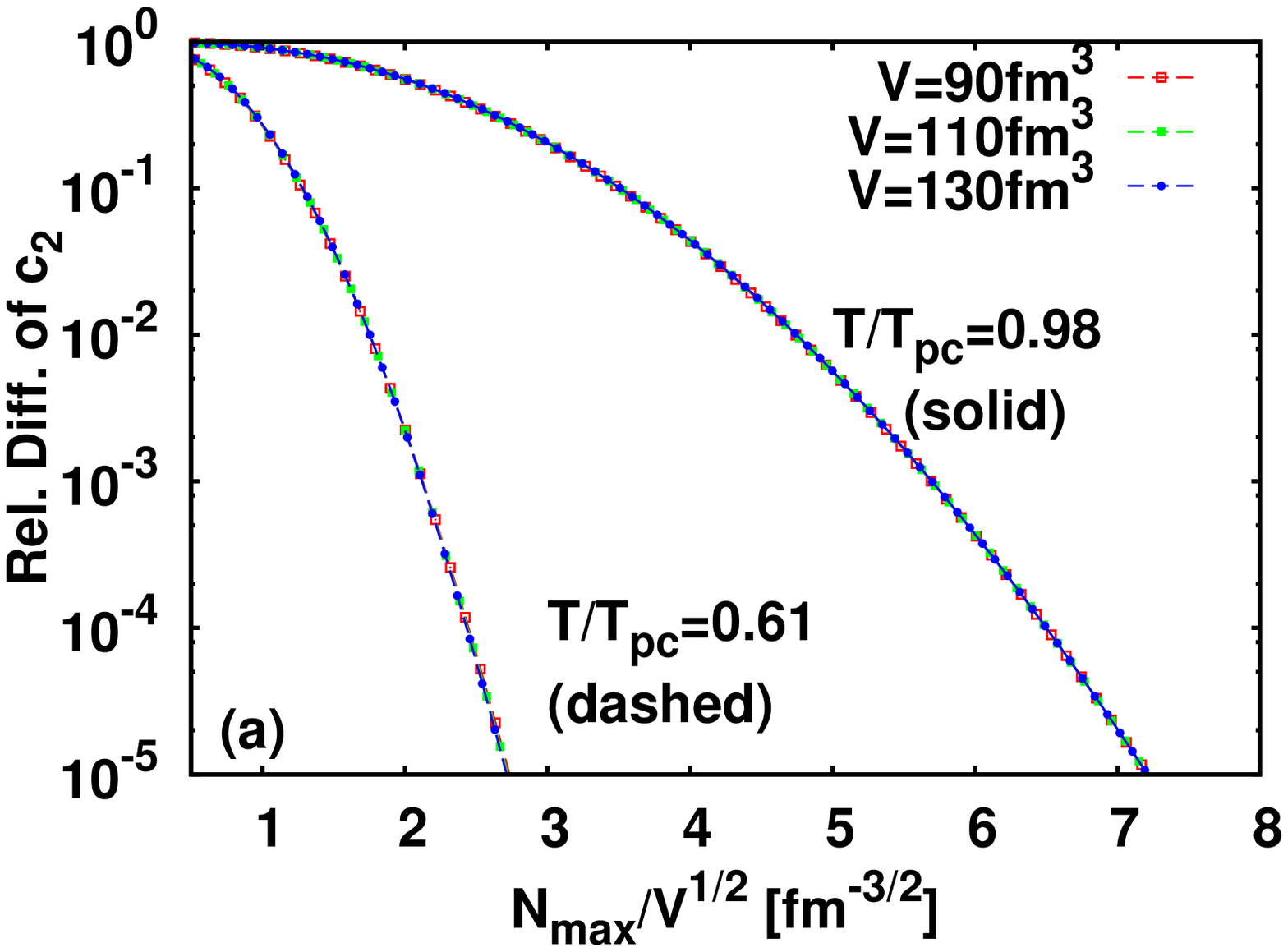}
 \includegraphics[width=0.32\textwidth]{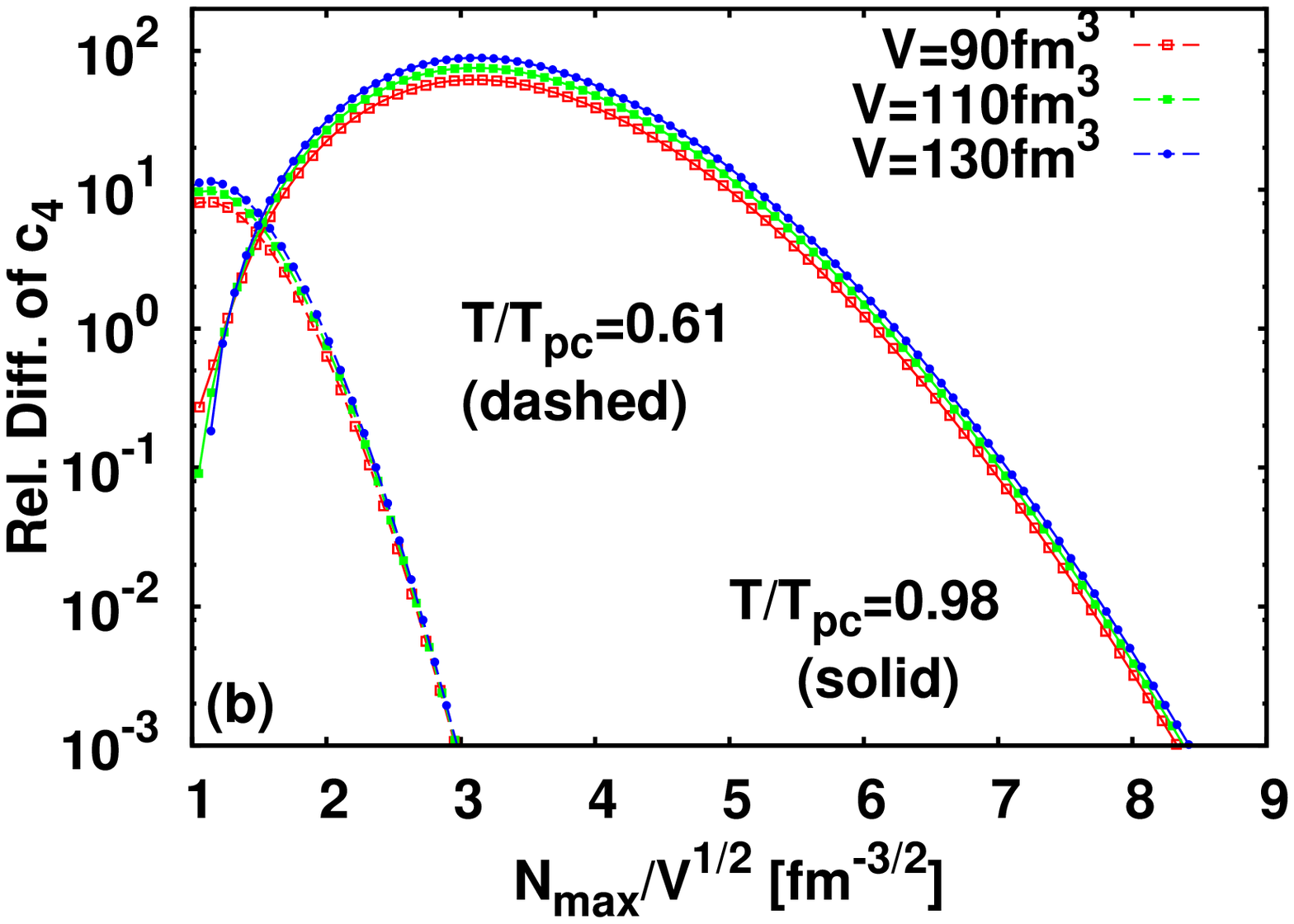}
 \includegraphics[width=0.32\textwidth]{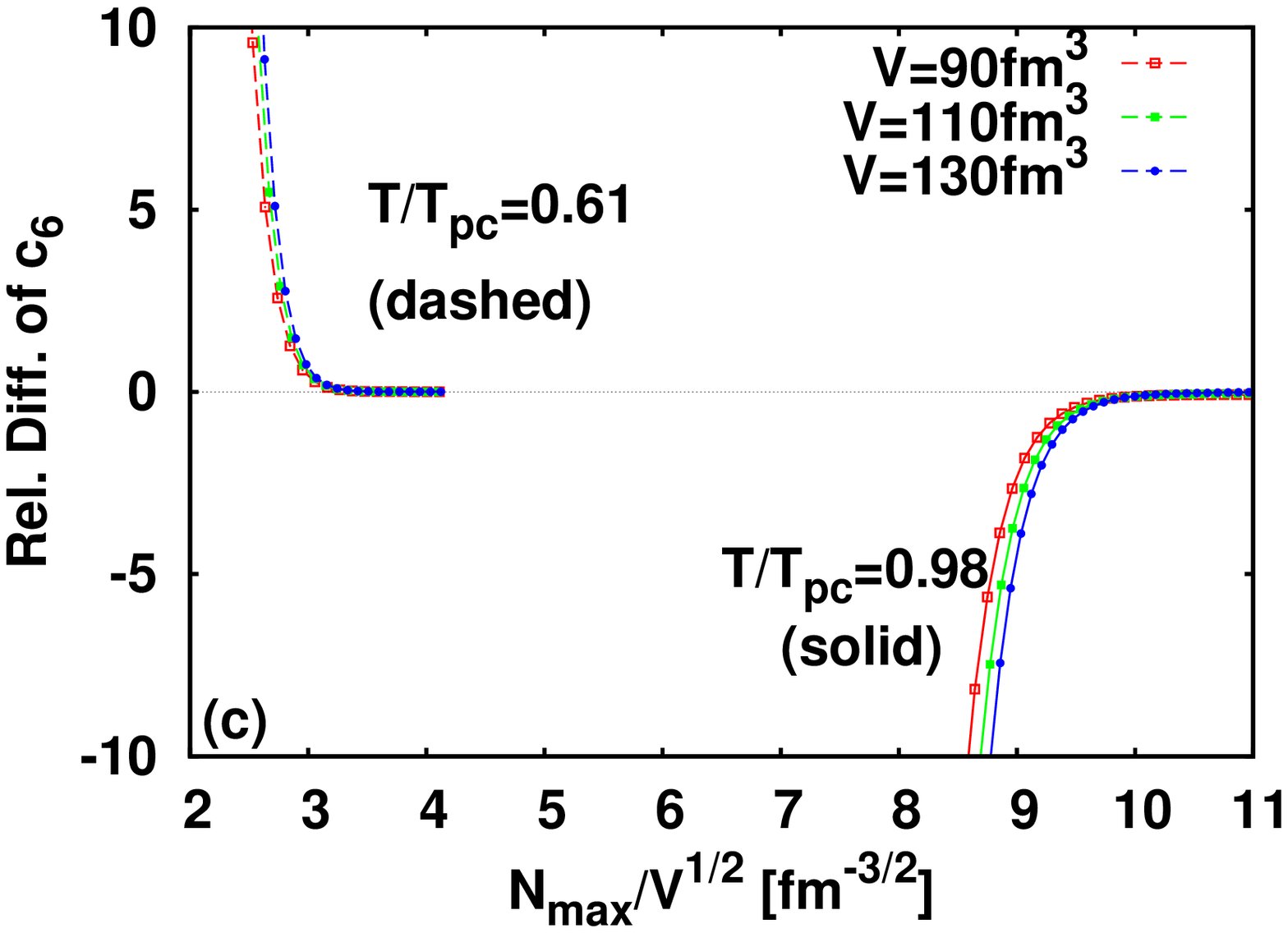}
 \caption{(Color online) Relative difference between cumulants $c_n$
 $(n=2,4,6)$  calculated in the quark-meson model within the FRG
 approach  using Eq.~\eqref{cumulants} and
 Eqs.~(\ref{nmax}-\ref{moments}) for two temperatures and for several
 values of the  volume parameters $V$ (see text).}
  \label{mom2}
\end{figure*}

  \subsection{Quantum statistics effects}

The Skellam distribution \eqref{eq:pn_skellam}, describes the fluctuations in
a gas of non-interacting classical particles with a conserved
charge. \textit{i.e.,} particles and antiparticles obeying the Poisson
distribution.

In general, effects of quantum statistics should be included if the mass
of the particle is smaller than the temperature.
The quantum statistics effects were  also shown to modify significantly
the probability distribution of the electric charge and the
resulting cumulants \cite{P_n_charge,Skokov:2011rq}.

 In the FRG approach to the quark-meson model,
and at $T/T_{pc}=0.98$ one finds,  that $M_q/T \simeq 0.6$.  Thus, the Boltzmann approximation is
clearly not justified  near $T_{pc}$.


In order to disentangle the effects of mesonic fluctuations and quantum
statistics, we also consider  the probability distribution
$P(N)$,  obtained from Eq.~\eqref{eq:cano_theta},   for a free gas of quarks and anti-quarks
 with the thermodynamic potential density
\begin{align}
 \Omega_{\text{Fermi}}(T,\theta)& = -\frac{\nu_q
  T}{2\pi^2}\int_{0}^{\infty}dk k^2 [\ln(1+e^{-E_k/T+i\theta}) \nonumber\\
 & \quad +\ln(1+e^{-E_k/T-i\theta})]\label{new},
\end{align}
and the  quark energy $E_k$ from
Eq.~\eqref{eq:skellam_baryonnum}.

The effect of Fermi statistics is illustrated in Fig.~\ref{fig:fermi}
with the dynamical quark mass obtained in the FRG approach. The
probability
distribution of the free Fermi gas is seen to be narrower
than the corresponding Skellam function, but still broader than the $P(N)$
of the quark-meson model in the FRG approach. We identify the residual
effect, i.e.\ the difference between the Fermi gas and the FRG
distributions, as being due to  the mesonic fluctuations implying
the $O(4)$ criticality near the chiral crossover transition.

On the other hand, in the MF approach at $T/T_{pc}=0.98$, the
net-quark probability distribution was found to be  slightly
broader  than the corresponding distribution obtained for a  free Fermi
gas. Therefore,  in the MF approach,  the apparently  narrower
distribution  than the Skellam, is due to the quantum statistics
effects. Thus, there is a clear difference in the properties  of the
net-quark distributions obtained under MF and the FRG approach. The
mesonic fluctuations result in a shrinking of the
distribution,  whereas the MF critical dynamics results in a broadening
of the distribution, relatively  to the non-singular Fermi gas
reference, as shown in \cite{morita12:_baryon_number_probab_distr_near}.

\subsection{Cumulants of the net quark number}

Fluctuations of the net-quark number are quantified by the corresponding cumulants $c_n(T,\mu)$,  which in turn reflect
critical fluctuations. Consequently, cumulants can be used to probe the phase diagram of QCD \cite{ejiri06:_hadron_fluct_at_qcd_phase_trans,karsch11:_probin_freez_out_condit_in,braun-munzinger11:_net_proton_probab_distr_in,stephanov09:_non_gauss_fluct_near_qcd_critic_point,stephanov11:_sign_of_kurtos_near_qcd_critic_point,skokov10:_vacuum_fluct_and_therm_of_chiral_model,friman11:_fluct_as_probe_of_qcd,skokov11:_quark_number_fluct_in_polyak}.

In statistical physics, the cumulants are related to the corresponding susceptibilities,
\begin{equation}
 c_n(T,\mu) \equiv \frac{\partial^n [p(T,\mu)/T^4]}{\partial (\mu/T)^n}. \label{cumulants}
\end{equation}
Thus,
given the thermodynamic pressure $p(T,\mu,V) = (T/V)\ln\mathcal{Z}$
in the grand-canonical ensemble,
the  cumulants $c_n(T,\mu)$ can be obtained by taking
derivatives of the thermodynamic pressure with respect to the chemical
potential.


The cumulants in Eq.~\eqref{cumulants} can be also obtained from
the probability distribution $P(N)$, 
through the central moments,
$\langle (\delta N)^k \rangle=  \langle(N-\langle N \rangle)^k\rangle$,  where
\begin{equation}
 \langle N^k \rangle = \sum_{N=-N_{\text{max}}}^{N_{\text{max}}} N^k P(N),  \label{nmax}
\end{equation}
and $c_n$  are  given then as polynomials in $\langle (\delta N)^k \rangle$.
At vanishing chemical potential,
the  first three non-vanishing cumulants read,
\begin{align}
 c_2&= \frac{\langle (\delta N)^2 \rangle}{VT^3} ,\\
 c_4&= \frac{\langle (\delta N)^4 \rangle - 3 \langle (\delta N)^2 \rangle^{2}}{VT^3} ,\\
 c_6&= \left[ \langle (\delta N)^6 \rangle -15 \langle (\delta N)^4 \rangle
 \langle (\delta N)^2 \rangle\right. \nonumber\\
  &\left.- 10 \langle (\delta N)^3 \rangle^{2} +30\langle (\delta N)^2 \rangle^{3}\right]/(VT^3)\label{moments} .
\end{align}

In principle, $N_{\text{max}}$ in Eq.~\eqref{nmax} is infinite.
In practice, however,  $N_{\text{max}}$ is always finite. Thus, the
cumulants obtained from Eqs.~\eqref{nmax}-\eqref{moments} are, in most
cases, approximations to the exact results obtained from
Eq.~\eqref{cumulants}.

Figure~\ref{fig:comps} shows different ratios of cumulants for the
Skellam distribution, calculated for fixed mean number of quarks
and antiquarks,  from
Eqs.~\eqref{nmax}-\eqref{moments} for different $N_{\text{max}}$.
%
Clearly,  to reproduce exact results $c_{2n}=(b+\bar{b})/VT^3$, one needs  $P(N)$  for a sufficiently
large $N=N_{\text{max}}$. This value increases with the order of the cumulant,  and  also depends on  the volume,  temperature and the chemical potential.

Figure \ref{mom24} shows  different  cumulants  obtained in the quark-meson model within the FRG approach
from Eq.~\eqref{cumulants} and  their  approximate values computed from
Eqs.~\eqref{nmax}-\eqref{moments}  as functions of
temperature,  for several values of $N_{\text{max}}$.
For comparison, the cumulants obtained from the corresponding Skellam
distribution are also shown in this figure. One notes,  that  $c_{2}$ and $c_{4}$ of the Skellam
and FRG distributions differ, while $c_{6}$ agrees for temperatures well
below $T_{pc}$. This behavior can be linked  to  the $\mu$--dependence
of a dynamical quark mass, which at $T \ll T_{pc}$ saturates as the
fourth order polynomial in $\mu$,  which  is not included
in the calculations of $c_n$ from the Skellam function.

The convergence properties  of the cumulants with $N_{\text{max}}$  in the
quark-meson model are similar to those found for the Skellam distribution.
The value of $N_{\text{max}}$ needed to obtain a good
approximation, grows with the order of
the cumulant. This reflects the fact,  that cumulants of
higher order are more sensitive to  tail of the
distribution. For the parameter set used in  Fig.~\ref{mom24},
all  cumulants up to the sixth order are well reproduced with $N_{\text{max}}\simeq 90$.
This also confirms consistency of the calculation of $P(N)$ within the quark-meson model.

 At $\mu=0$, the second and the fourth
order cumulants  are not influenced  by the critical chiral dynamics, since
they remain finite,  even in the chiral limit. Thus, their properties are entirely
determined by the regular part of the partition function. The temperature
dependence of $c_2$ and $c_4$,  seen in Fig.~\ref{mom24}, is essentially
that of an ideal  quark  gas,  with a modified dispersion relation  by
$T$-- and $\mu$--dependence of a   dynamical quark  mass.

In  contrast, the temperature dependence of $c_6$, and in particular its negative values near $T_{pc}$, seen in Fig.~\ref{mom24}, are
 universal. The characteristic shape of $c_6$ is generic for
the $O(4)$ universality class, owing to the form of the
scaling function \cite{karsch11:_probin_freez_out_condit_in}.
The sixth order cumulant obtained from the non-critical Skellam
distribution, exhibits a very different behavior.

It is interesting to note, that already for moderate values of
$N_{\text{max}}$, the $O(4)$  shape of $c_6$  is qualitatively
reproduced from the probability distribution.
%
%
This result is of interest for the event-by-event analysis of heavy-ion collisions,
where one expects to see the $O(4)$ criticality by reconstructing
the  higher order moments from the measured net-charge probability distribution
\cite{aggarwal10:_higher_momen_of_net_proton}.

The deviations of the approximate cumulants $c_n^{A}$, obtained from  the probability distribution
\eqref{nmax}-\eqref{moments}, from their exact values  $c_n^{E}$,
given by Eq.~\eqref{cumulants}, depend on the volume of the system.
The relative deviations of $c_n^{A}$ from
$c_n^{E}$ at fixed $T$ for different $V$, however, obey the approximate
scaling relations.
This is transparent from Fig.~\ref{mom2},  showing the relative difference,
$R=(c_n^{E}-c_n^{A})/c_n^{E}$ for $n=2,4,6$,
as a function of $N_{\text{max}}/\sqrt{V}$.
For $n=2$ there is a clear  scaling for all $N_{\text{max}}$.  For higher
order cumulants,  the approximate scaling becomes better for
larger values of $N_{\text{max}}/\sqrt{V}$.

 \begin{figure}[!t]
  \includegraphics[width=3.375in]{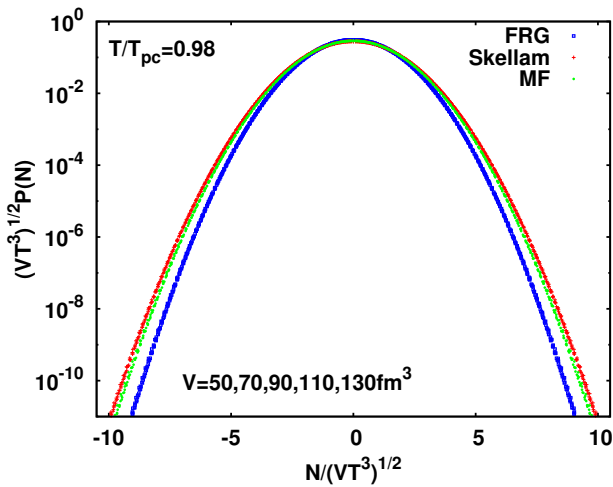}
  \caption{(Color online) Volume scaling of the net quark number probability distribution at fixed
  $T/T_{pc}$ in the quark-meson model within the MF,  the FRG approach  and for the
  corresponding Skellam distribution. }
  \label{mom46}
 \end{figure}

Since the cumulants of the net quark number are linked to the corresponding
probability distribution through Eqs.~\eqref{nmax}-\eqref{moments}, the
approximate scaling of the
cumulants, seen  in  Fig.~\ref{mom2},  should be also  reflected
in  the net quark number probability distributions.
Indeed,  Figure~\ref{mom46}  shows,  that for fixed $T$, the
$\sqrt{VT^3} P(N)$  scales approximately with
$N/\sqrt{VT^3}$. This  approximate scaling is valid for
$P(N)$, calculated  in the quark-meson
model  within the FRG and MF approach, as well as,  for the  corresponding Skellam
distributions. Figure~\ref{mom46} indicates, that the shape of the
distribution reflects the underlying criticality in a system  which is
governed by  the critical exponents. The probability distribution in the
MF case is broader than in the FRG approach, owing to differences
caused by the mesonic fluctuations. As they smoothen the
transition, the resultant dynamical quark mass is heavier than the MF
case at the same $T/T_{pc}$ \cite{skokov10:_meson_fluct_and_therm_of}, leading to narrower distribution.
%

\section{Concluding remarks}\label{sec:summary}

We have studied the properties of the probability distribution $P(N)$
of the net quark number  in the presence of the critical chiral
dynamics governed by the $O(4)$ universality class in the
quark-meson  model.  The  computations of $P(N)$ have been  done  within
the Functional Renormalization Group (FRG) approach, which preserves the
$O(4)$ scaling of relevant physical observables.

The main objectives of this paper  was to study the influence of the
underlying chiral phase transition on the net quark probability distribution,
for a physical value of the pion mass.

 We have shown, by comparing the FRG and  the mean-field (MF) results,
that the shape of the distribution reflects the criticality in a system.
The FRG distribution is narrower than the one obtained in the  MF
approximation. This is mainly due to  differences in {the} values of
dynamically generated   quark mass. Effects of the expected $O(4)$
criticality appear in the tail of the distribution and imply
characteristic shapes of the higher order cumulants. 

Near the chiral crossover transition, the
probability distribution of the net quark number was also shown to be
narrower than the Skellam function, which corresponds to a classical
quasiparticle gas and is used as a reference for the
non-critical distribution.
The narrowing of the probability distribution is mainly due to  mesonic
fluctuations.  However, owing to dropping quark mass near the chiral
transition also quantum statistics play a role, although a sub-leading
one.

The observed shrinking of $P(N)$ relative to the Skellam distribution is
also expected in the $O(2)$ universality class.
This is because, the scaling functions in the $O(4)$ and $O(2)$ universality
classes are very similar
\cite{ejiri09:_magnet_equat_of_state_in_flavor_qcd} and exhibit negative
values of the specific heat critical exponents $\alpha$. However, since
$\alpha$ in $O(2)$ is larger than in $O(4)$ universality class, one
expects quantitative differences in the properties of $P(N)$.

We have  found  an approximate  scaling  of the probability
distribution and of the net charge cumulants with the volume of the
system. This implies,
that the observed properties  of $P(N)$  near the chiral transition are
volume independent,  and are due to mesonic fluctuations implying
 $O(4)$ criticality near the chiral crossover transition.

{These} results are  of importance in heavy ion
collisions,  where by measuring the net baryon number probability
distribution  and related  moments,  one expects to {experimentally probe} the QCD
phase boundary. { The phenomenological implications of our results will
be presented elsewhere  \cite{paper3}.}

\acknowledgements
 The authors acknowledge stimulating discussions with
 P. Braun-Munzinger, F. Karsch and Nu Xu.
 K.M. was supported by Yukawa International Program for
 Quark-Hadron Sciences at  Kyoto University and by the Grant-in-Aid for
 Scientific Research from JSPS No.24540271. B.F. is supported in part
 by the Extreme Matter Institute EMMI.
 K.R. acknowledges partial support of the Polish Ministry of National
 Education (MEN).
 The research of V.S. is supported under Contract No. DE-AC02-98CH10886
 with the U. S. Department of Energy.

\end{document}